\documentclass{anstrans}
\title{An Accurate S$_N$ Method for Solving Static Multigroup Neutron Transport Equations in Slab Geometry}
\author{Jilang Miao,$^{*1}$ and Miaomiao Jin$^{*}$}

\institute{
$^{*}$Department of Nuclear Engineering, The Pennsylvania State University, 218 Hallowell Building, University Park, 16802 PA, USA
}
\footnotetext[1]{\footnotesize Corresponding author: Jilang Miao (jlmiao@psu.edu)}

\usepackage{graphicx} 
\usepackage{booktabs} 
\usepackage{subfigure}
\usepackage{microtype} 
\usepackage{upgreek}

\begin{document}
\section{Introduction }
This paper presents an accurate S$_N$\cite{Car1953}\cite{Lar2008} solver for slab geometry.
For constant cross-section regions, it gives accurate angular fluxes without need for fine meshes or approximation of solution forms.
The method provides a potentially accurate and efficient axial solver in the 2D-1D scheme~\cite{stimpson2015azimuthal}~\cite{hursin2014development} to solve 3D transport equations.

In this summary, we first derive the solution form for a constant cross-section region. 
The solution generalizes the earlier work~\cite{analytical2A2G} considering an 1D problem with only two angles and two groups to any number of energy groups and discrete angles. 
Then we show the steps to find the coefficients for each region from boundary conditions of the slab.
Finally, a two group case problem is studied with S$_2$,S$_4$,S$_6$ and results are verified with references from Monte Carlo simulations.

\section{Theory}
\subsection{S$_N$ equation in a homogeneous slab}
For energy groups $g=1,...,G$ and a quadrature set $\left . \{\mu_n,\omega_n\} \right | _{n=1,...,N}$, 
the transport equation for angular flux $\psi_{g,n}$ can be written as in Eq~\ref{eq::sn1d}.

\begin{equation}
\begin{aligned}
& \mu_n \frac{\partial}{\partial x} \psi_{g,n}(x)+  \Sigma_{t, g} \psi_{g,n}(x)= \\
&   \sum_{n^{\prime}, g^{\prime}} \omega_{n \prime} \Sigma_{s, g^{\prime} n^{\prime} \rightarrow g n} \psi_{g^{\prime}, n^{\prime}}(x) \\
&+\frac{1}{k_{eff}}  \sum_{n^{\prime}, g^{\prime}} \omega_{n^{\prime}}  \nu \Sigma_{f, g^{\prime} n^{\prime} \rightarrow g n} \psi_{g^{\prime}, n^{\prime}}(x) 
\label{eq::sn1d}
\end{aligned}
\end{equation}

The angular fluxes can be represented in a $N\times G$ vector $\Psi(x)$, such that $\psi_{g,n}(x)=\Psi_{g\times N + n}(x)$. 
The cross-sections and quadrature constants are defined in matrices below to facilitate writing Eq~\ref{eq::sn1d} in matrix form. 

Define diagonal matrix $\upmu$ with 
\begin{equation}
    \upmu_{n,n}= \frac{1}{\mu_n}. 
\end{equation}

Define diagonal matrix $\Omega$ with 
\begin{equation}
    \Omega_{n,n}= \omega_n. 
\end{equation}

$I_M$ is the $M\times M$ identity matrix.
$1_M$ is an $N\times N$ matrix with all elements being 1.

Define diagonal matrix $T$ from total cross-section as 
\begin{equation}
    T_{xs,gN+n,gN+n}=\Sigma_{t,g}. 
\end{equation}

Define matrix $F_{xs}$ from fission cross-section as 
\begin{equation}
    F_{xs,g'N+n',gN+n}=\nu \Sigma_{f, g^{\prime} n^{\prime} \rightarrow g n}. 
\label{eq::F}
\end{equation}
In Eq~\ref{eq::F}, 
$\nu \Sigma_{f, g^{\prime} n^{\prime} \rightarrow g n}$ describes general  source-dependent angle and energy distribution of neutrons out of fission.
If fission is not source dependent and isotropic, $\nu \Sigma_{f, g^{\prime} n^{\prime} \rightarrow g n}$ can be simplified to 
\begin{equation}
    \nu \Sigma_{f, g^{\prime} n^{\prime} \rightarrow g n} 
    = \frac{1}{ \sum_{n^{\prime}} w_{n'}} \chi_g \nu\Sigma_{f,g'}
\label{eq::nuSigf}
\end{equation}

Define matrix $S_{xs}$ from fission cross-section as 
\begin{equation}
    S_{xs,g'N+n',gN+n}=\Sigma_{s, g^{\prime} n^{\prime} \rightarrow g n}. 
\label{eq::F}
\end{equation}
If scattering is assumed to be isotropic , 
\begin{equation}
    \Sigma_{s, g^{\prime} n^{\prime} \rightarrow g n} 
    = \frac{1}{ \sum_{n^{\prime}} w_{n'}} \Sigma_{s,g^{\prime} \rightarrow g }
\label{eq::SigS}
\end{equation}

Then an $NG\times NG$ Matrix $A$ are constructed from the cross-section matrices ($T_{xs}$,$F_{xs}$,$S_{xs}$  as below. 
\begin{align}
     T &= ( I_G\otimes \upmu ) T_{xs} \\
     F &= ( I_G\otimes \upmu ) T_{xs} ( I_G\otimes \Omega ) \\
     S &= ( I_G\otimes \upmu ) S_{xs} ( I_G\otimes \Omega ) \\
     A &= \frac{F}{k_{eff}} + S - T 
    \label{eq::TFStoA}
\end{align}
Finally, Eq~\ref{eq::sn1d} can be written in matrix form as in Eq~\ref{eq::sn1d matrix}.
\begin{equation}
    \partial_x \Psi(x) = A \Psi(x)
    \label{eq::sn1d matrix}
\end{equation}

\subsection{S$_N$ solution in a homogeneous slab}
Solution of Eq~\ref{eq::sn1d matrix} can be read after (block-) diagonalize~\cite{strang2006linear} matrix A as in Eq~\ref{eq::AP=PB}. 
\begin{equation}
A P = P B    
\label{eq::AP=PB}
\end{equation}

For the real matrix $A$, the eigenvalues are either real numbers or complex number in conjugate pairs. 
We choose all the real eigenvalues, and one complex eigenvalue of each conjugate pairs and denote them as 
\begin{equation}
\Lambda=\left[\begin{array}{cccc}
\lambda_1 & \lambda_2 & \ldots & \lambda_r
\end{array}\right]
\end{equation}
where $r$ is the number of eigenvalues selected after discarding one of each conjugate pair. 
And the transform matrix $P$ can be constructed from the corresponding eigenvalues as
\begin{equation}
P=\left[\begin{array}{cccc}
P_1 & P_2 & \ldots & P_r
\end{array}\right]
\end{equation}

If $\lambda_i$ is real, $P_i$ is the corresponding eigenvector $u_i$.
If $\lambda_j$ is complex, $P_j$ has two columns, the real and imaginary part of the corresponding eigenvector $u_j$:
\begin{equation}
P_j=\left[\begin{array}{cc}
\operatorname{Re}\left( u_j\right)  & 
\operatorname{Im}\left(u_j \right) 
\end{array}\right]
\end{equation}

The block-diagonal matrix $B$ is constructed from the $r$ eigenvalues as in Eq~\ref{eq::B form}.
\begin{equation}
\left[\begin{array}{cccc}
C_1 & 0 & \ldots & 0 \\
0 & C_2 & \ldots & 0 \\
\ldots & \ldots & \ldots & \ldots \\
0 & 0 & \ldots & C_r
\end{array}\right]
\label{eq::B form}
\end{equation}
If $\lambda_i$ is real, 
\begin{equation}
C_i=\lambda_i \equiv r_i
\end{equation}

If $\lambda_j$ is complex, 
\begin{equation}
C_j=\left[\begin{array}{cc}
\operatorname{Re}\left(\lambda_j\right) & \operatorname{Im}\left(\lambda_j\right) \\
-\operatorname{Im}\left(\lambda_j\right) & \operatorname{Re}\left(\lambda_j\right)
\end{array}\right]
\equiv \left[\begin{array}{cc}
r_j & m_j\\
-m_j & r_j
\end{array}\right]
\end{equation}

Since the $\partial_x$ operator commutes with a constant matrix ($P^{-1}$), the equation of $\Psi(x)$ can be transformed to 
\begin{equation}
    \partial _x X(x) + B X(x) = 0 
    \label{eq:sX + BX=0}
\end{equation}
where 
\begin{equation}
    X(x) = P^{-1} \Psi(x)
    \label{eq::X=invP Psi}
\end{equation}

With the matrix $B$ taking form as in Eq~\ref{eq::B form}, 
solution of Eq~\ref{eq:sX + BX=0} can be written as
\begin{equation}
    X(x) = \Gamma(x) \alpha
    \label{eq::X=G a}
\end{equation}
$\Gamma(x)$ has the form
\begin{equation}
\left[\begin{array}{cccc}
\Gamma_1(x) & 0 & \ldots & 0 \\
0 & \Gamma_2(x) & \ldots & 0 \\
\ldots & \ldots & \ldots & \ldots \\
0 & 0 & \ldots & \Gamma_r(x)
\end{array}\right]
\label{eq::G form}
\end{equation}

If $\lambda_i$ is real, 
\begin{equation}
\Gamma_i(x)=e^{r_i x}
\end{equation}

If $\lambda_j$ is complex, 
\begin{equation}
\Gamma_j(x)=\mathrm{e}^{r_j x} \left[\begin{array}{cc}
 \mathrm{cos}(m_j x) & \mathrm{sin}(m_j x)\\
-\mathrm{sin}(m_j x) & \mathrm{cos}(m_j x)
\end{array}\right]
\end{equation}

$\alpha$ in Eq~\ref{eq::X=G a} is to be found from boundary conditions. 

\subsection{S$_N$ solution in slab with heterogeneous regions}
Consider a slab system with $R$ homogeneous regions.
Number the regions as $1,...,R$ from left to right. 
And the position of the $R+1$ surfaces are defined as $x_0,...,x_R$.
The S$_N$ solution requires the $\alpha$ coefficients (Eq~\ref{eq::X=G a}) for each region. 
The coefficients can be solved from the boundary conditions on the left and right ends and continuity of the angular flux on region interfaces. 

\subsubsection{Incoming source boundary condition}
For incoming source $\Psi_L$ from the left end (angular flux for the $N/2$ angles with $\mu>0$), the boundary condition can be represented as
\begin{equation}
    \left. \left( P_{1} \Gamma_{1} (x_0) \right)\right| _{\left\{\mu>0\right\}} \alpha_1 = \Psi_L
    \label{eq::BC srcL}
\end{equation}

Similarly,for incoming source $\Psi_R$ from the right end, the boundary condition can be represented as
\begin{equation}
    \left. \left( P_{R} \Gamma_{R} (x_R) \right)\right| _{\left\{\mu<0\right\}} \alpha_R = \Psi_R
\label{eq::BC srcR}
\end{equation}

\subsubsection{Reflective boundary condition}
The reflective boundary can be represented as Eq~\ref{eq::BC ref},  
\begin{equation}
   \left[
   \left.\left( P_{r} \Gamma_{r} (x_i) \right)\right|_{\left\{\mu<0\right\}} -
   \left.\left( P_{r} \Gamma_{r} (x_i) \right)\right|_{\left\{\mu>0\right\}}
   \right]\alpha_r = 0  
   \label{eq::BC ref}
\end{equation}
where $r=1,R$, and $x_i=x_0$ if $r=1$ and $x_i=x_R$ if $r=R$. 

\subsubsection{Angular flux continuity condition}
At region interfaces, all angular fluxes are continuous on both sides.
The condition for the surface between region $i$ and $i+1$ is represented as 
\begin{equation}
    P_{i} \Gamma_{i} (x_i) \alpha_i  -
     P_{i+1} \Gamma_{i+1} (x_i)   \alpha_{i+1}
    = 0  
   \label{eq::BC cont}
\end{equation}

\subsubsection{Solution of the coefficients for each region}
For boundary cells,each of Eq~\ref{eq::BC srcL}~\ref{eq::BC srcR}~\ref{eq::BC ref} only provides $NG/2$ rows (equations).
For interior cells, each of Eq~\ref{eq::BC cont} provides $NG$ rows.
And they can form $NG\times R$ to solve the $NG\times R$ coefficients in $\left. \left\{\alpha_i\right\} \right|_{i=1,...,R}$. 

If both ends have incoming source, $\left. \left\{\alpha_i\right\} \right|_{i=1,...,R}$ can be solved from Eq~\ref{eq::BCmat VV}. 

\begingroup
\scriptsize
\begin{equation}
\begin{split}
\left[\begin{array}{cccc}
\left(\begin{array}{c} 
\left. \left( P_1 \Gamma_1(x_0) \right)\right|_{\left\{\mu>0\right\}} \\ 0 
\end{array}
\right)
& 0 & \ldots  & 
\left(\begin{array}{c} 
0 \\ 
\left. \left( P_R \Gamma_R(x_R) \right)\right|_{\left\{\mu<0\right\}} 
\end{array}
\right)
\\
P_1 \Gamma_1(x_1) & P_2 \Gamma_2(x_1) & \ldots &  0 \\
\ldots & \ldots & \ldots &   \ldots \\
0 &  \ldots & P_{R-1}\Gamma_{R-1}(x_R) & P_R\Gamma_R(x_R)
\end{array}\right] \alpha \\ = 
\left[\begin{array}{c}
\left(
\begin{array}{c} 
\Psi_L \\ \Psi_R
\end{array}
\right)  
 \\
0 \\
0 \\
0
\end{array}\right]
\label{eq::BCmat VV}
\end{split}
\end{equation}
\endgroup

If there is incoming source from right end and left end is reflective, $\left. \left\{\alpha_i\right\} \right|_{i=1,...,R}$ can be solved from Eq~\ref{eq::BCmat RV}. 
\begingroup
\scriptsize
\begin{equation}
\begin{split}
\left[\begin{array}{cccc}
\left(\begin{array}{c} 
\left. \left( P_1 \Gamma_1(x_0) \right)\right|_{\left\{\mu<0\right\}} -
\left. \left( P_1 \Gamma_1(x_0) \right)\right|_{\left\{\mu>0\right\}}
\\ 0 
\end{array}
\right)
& 0 & \ldots  & 
\left(\begin{array}{c} 
0 \\ 
\left. \left( P_R \Gamma_R(x_R) \right)\right|_{\left\{\mu<0\right\}} 
\end{array}
\right)
\\
P_1 \Gamma_1(x_1) & P_2 \Gamma_2(x_1) & \ldots &  0 \\
\ldots & \ldots & \ldots &   \ldots \\
0 &  \ldots & P_{R-1}\Gamma_{R-1}(x_R) & P_R\Gamma_R(x_R)
\end{array}\right] \alpha \\ = 
\left[\begin{array}{c}
\left(
\begin{array}{c} 
0 \\ \Psi_R
\end{array}
\right)  
 \\
0 \\
0 \\
0
\end{array}\right]
\label{eq::BCmat RV}
\end{split}
\end{equation}
\endgroup
For an eigenvalue problem,
Eq~\ref{eq::BCmat VV} and Eq~\ref{eq::BCmat RV} have a zero right hand side,
and $k_{eff}$ can be found by making the determinant on the left hand being 0. 
\section{Results and Analysis}
In this section, case studies are performed on a slab system . 
The reactor core is located within $[-\frac{b}{2},\frac{b}{2}]$ .
And reflector areas are within $[-\frac{a}{2},-\frac{b}{2}]$ and $[\frac{b}{2},\frac{a}{2}]$.  
The geometry is set up with $a=0.5$ and $b=0.4$.

Two-group cross-sections for the core and reflector materials are shown in Table~\ref{tab::xs}. 
Isotropic scattering and fission neutrons are assumed.
They are also assumed to be source-independent. 
\begin{table}[htb]
  \centering
  \caption{cross-section parameters}
  \begin{tabular}{llr}\toprule
      &  core      & reflector
\\ \midrule
$\Sigma_{t,1}$              & 6.667 &13.333 \\
$\Sigma_{t,2}$              & 8.333 &16.667 \\
$\Sigma_{s,1\rightarrow 1}$ & 2     & 4     \\
$\Sigma_{s,1\rightarrow 2}$ & 3.333 & 8     \\
$\Sigma_{s,2\rightarrow 1}$ & 0     & 0.833 \\
$\Sigma_{s,2\rightarrow 2}$ & 4.583 & 13.333 \\
$\Sigma_{f,1}$              & 0     & 0.04   \\
$\Sigma_{f,2}$              & 2.917 & 0.067 \\
$\nu$                       & 2     & 0 \\
$\chi_1$                    & 1     & 0      \\
$\chi_2$                    & 0     & 0     \\
\bottomrule
\end{tabular}
  \label{tab::xs}
\end{table}

A reference solution is generated using OpenMC~\cite{romano2013openmc} multigroup mode with the same cross-sections as in Table~\ref{tab::xs}. 
The slab has vacuum boundary on $x=\pm a$, and  the reflective $y$ and $z$ planes are set up at $\pm 100$. 
The simulation tracks $10^6$ neutrons per generation.
The neutrons are simulated for $200$ generations and tallies started for the next $1000$ generations. 
Total flux and $k_{eff}$ were tallied. 

The S$_N$ solver used reflective boundary at the center and vacuum boundary at the right end. 
Eq~\ref{eq::BCmat RV} is solved for $N=2,4,6$. 
Gauss–Legendre quadrature sets are used. 
The eigensystem of matrix $A$(Eq~\ref{eq::TFStoA}) for reach region is solved using the Numpy linear algebra package~\cite{harris2020array}.
The complexity of eigenvalue decomposition is $\mathcal{O}(NG^3)$ for each of the $R$ meshes. 
A basic bisection routine is used to search $k_{eff}$ that makes the determinant of Eq~\ref{eq::BCmat RV} be zero. 

Table~\ref{tab::keff} shows the $k_{eff}$ from openMC and the different orders of S$_N$ solvers.
It clearly shows how higher order solution approaches the Monte Carlo reference. 

\begin{table}[htb]
  \centering
  \caption{$k_{eff}$ of different S$_N$ orders compared with Monte Carlo}
  \begin{tabular}{llr}\toprule
Method      & $k_{eff}$      & $k_{eff}$ - $k_{keff,MC}$ (pcm) 
\\ \midrule
Monte Carlo  & 0.96606 $\pm$ 3$\times 10^{-5}$ &  \\
$S_2$        & 0.95979 & -627 \\
$S_4$        & 0.96612 &   6.4 \\
$S_6$        & 0.96609 &   2.9 
\\
\bottomrule
\end{tabular}
  \label{tab::keff}
\end{table}

\begin{figure}[htbp]
\centering
\subfigure
{\includegraphics[width=0.475\textwidth]{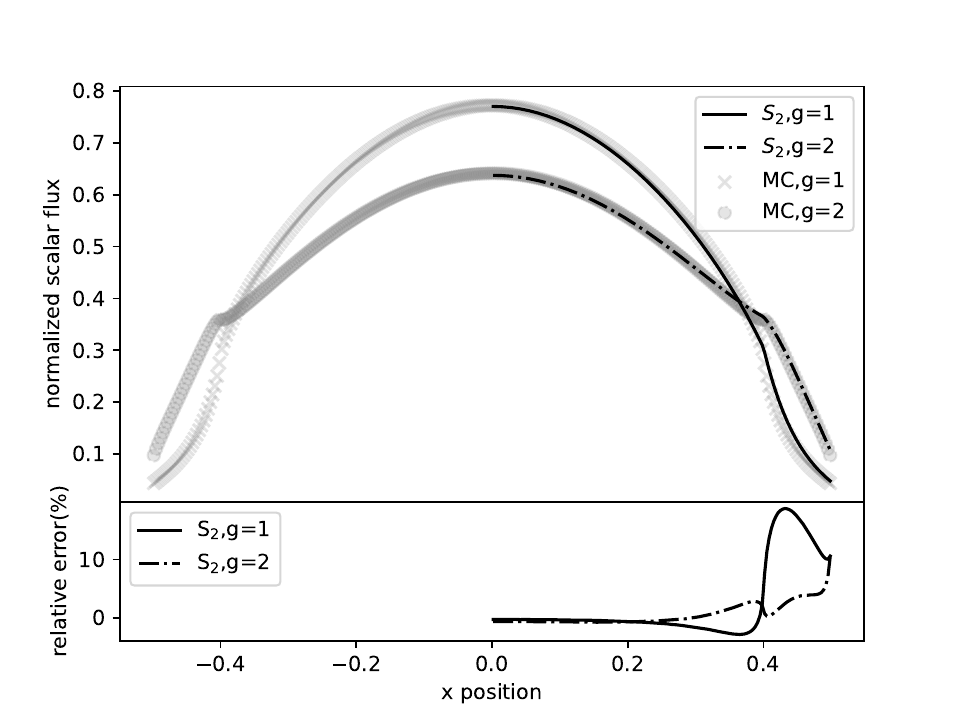}
\label{fig::phiS2}} 
\subfigure
{\includegraphics[width=0.475\textwidth]{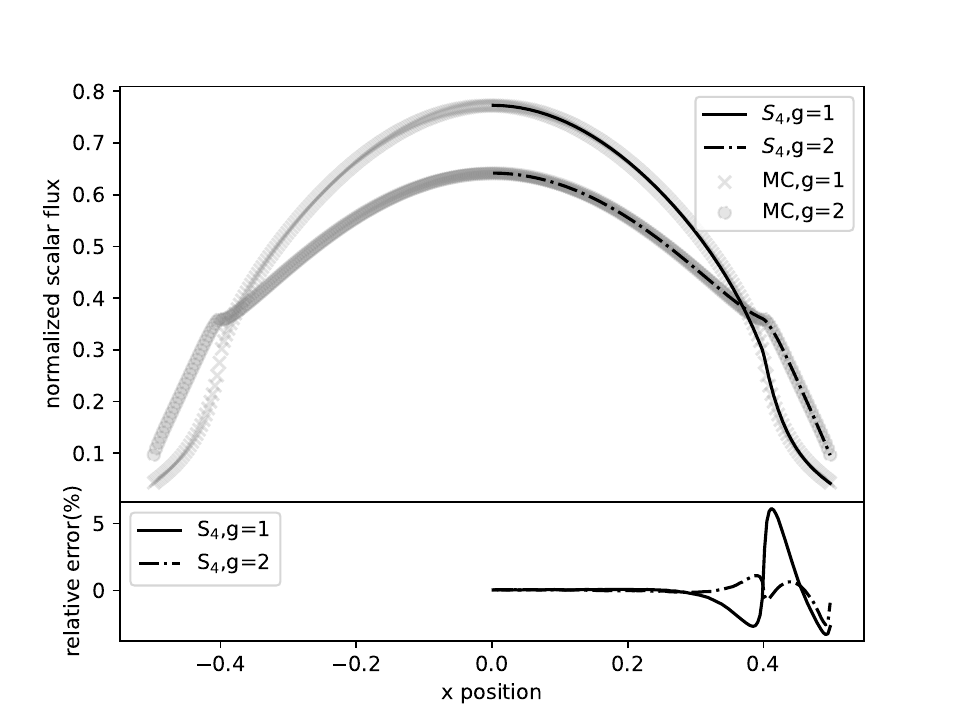}
\label{fig::phiS4}}  
\subfigure
{\includegraphics[width=0.475\textwidth]{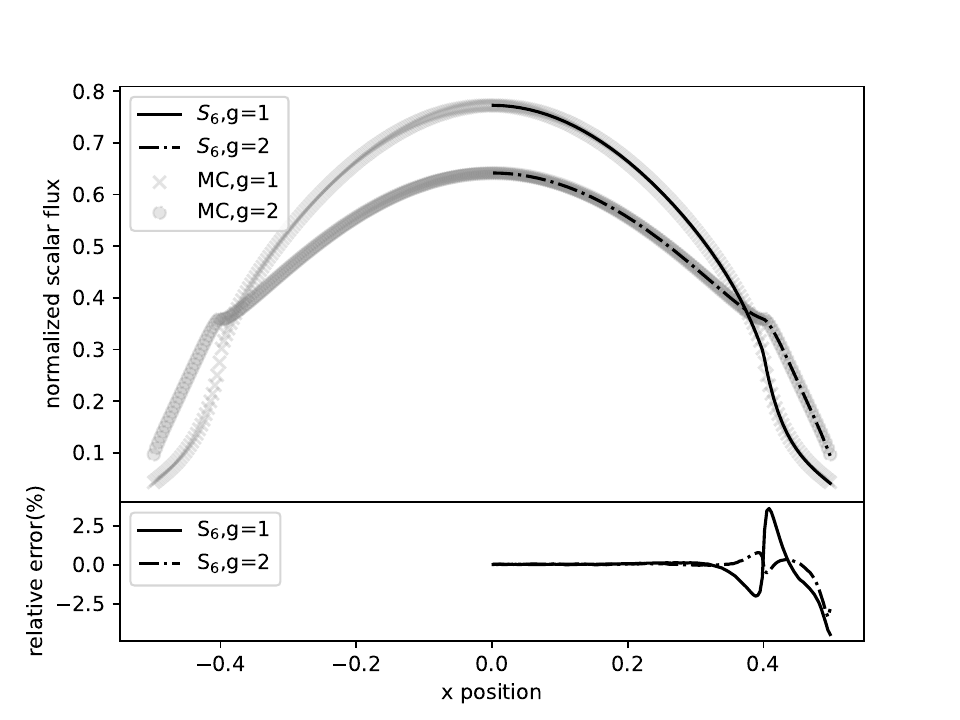}
\label{fig::phiS6}}  
\caption{Scalar fluxes from $S_N$ vs Monte Carlo }
\label{fig::phi}
\end{figure}

Fig~\ref{fig::phi} shows the normalized scalar fluxes. 
The relative error between S$_N$ and reference at each position is also plotted below the scalar fluxes. 
In Fig~\ref{fig::phiS2},$S_2$ gives roughly right shape of scalar fluxes. 
The largest difference occurs at the interface with around $20\%$ relative error. 
Better performance of higher order solutions is observed. 
S$_4$ in Fig~\ref{fig::phiS4} shows better matched scalar flux shapes and the maximum relative error decreases to around $5\%$. 
S$_6$ in Fig~\ref{fig::phiS6}) reaches below $2.5\%$ relative error. 

\begin{figure}[htbp]
\centering
\subfigure
{\includegraphics[width=0.475\textwidth]{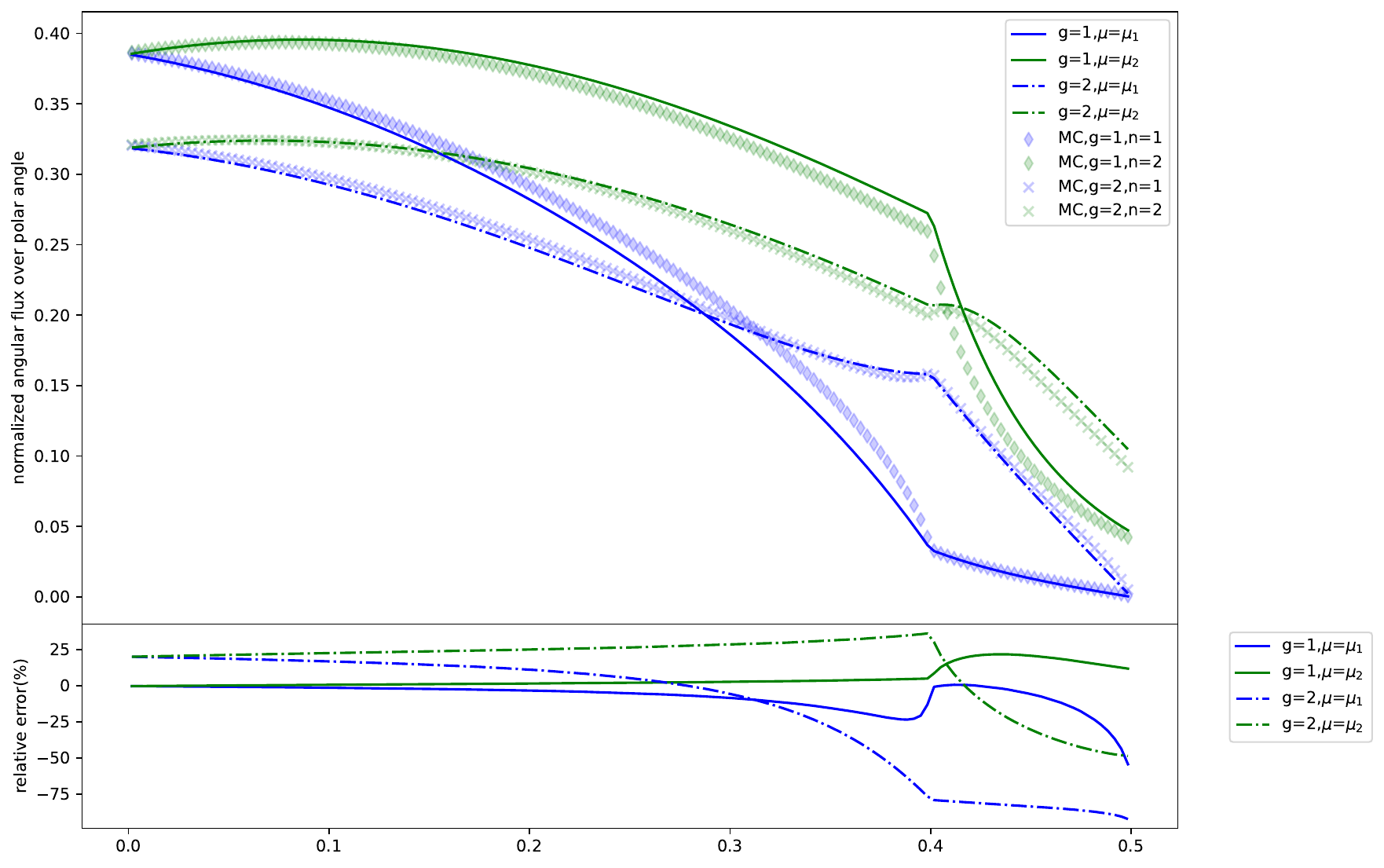}
\label{fig::psiS2}} 
\subfigure
{\includegraphics[width=0.475\textwidth]{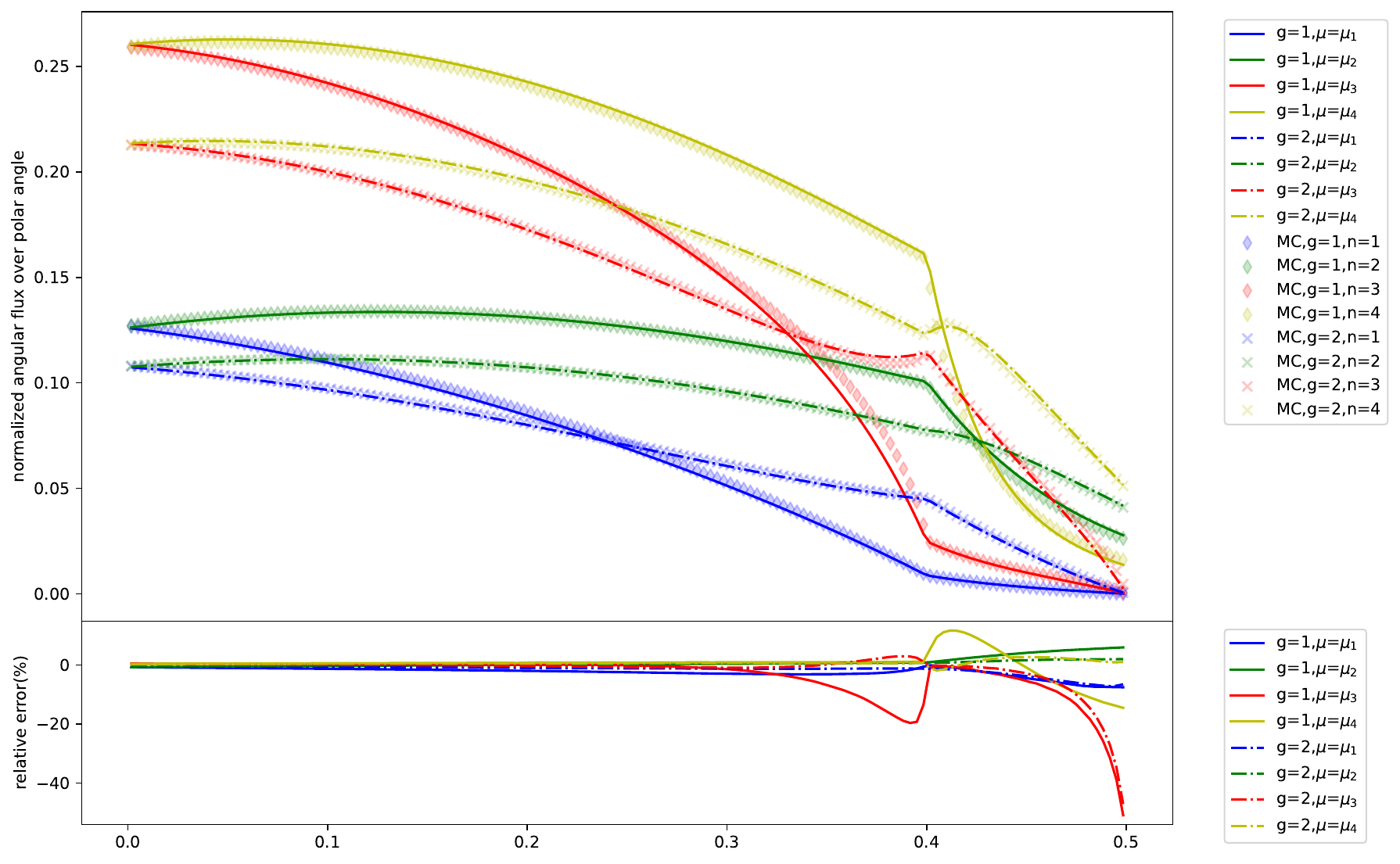}
\label{fig::psiS4}}  
\subfigure
{\includegraphics[width=0.475\textwidth]{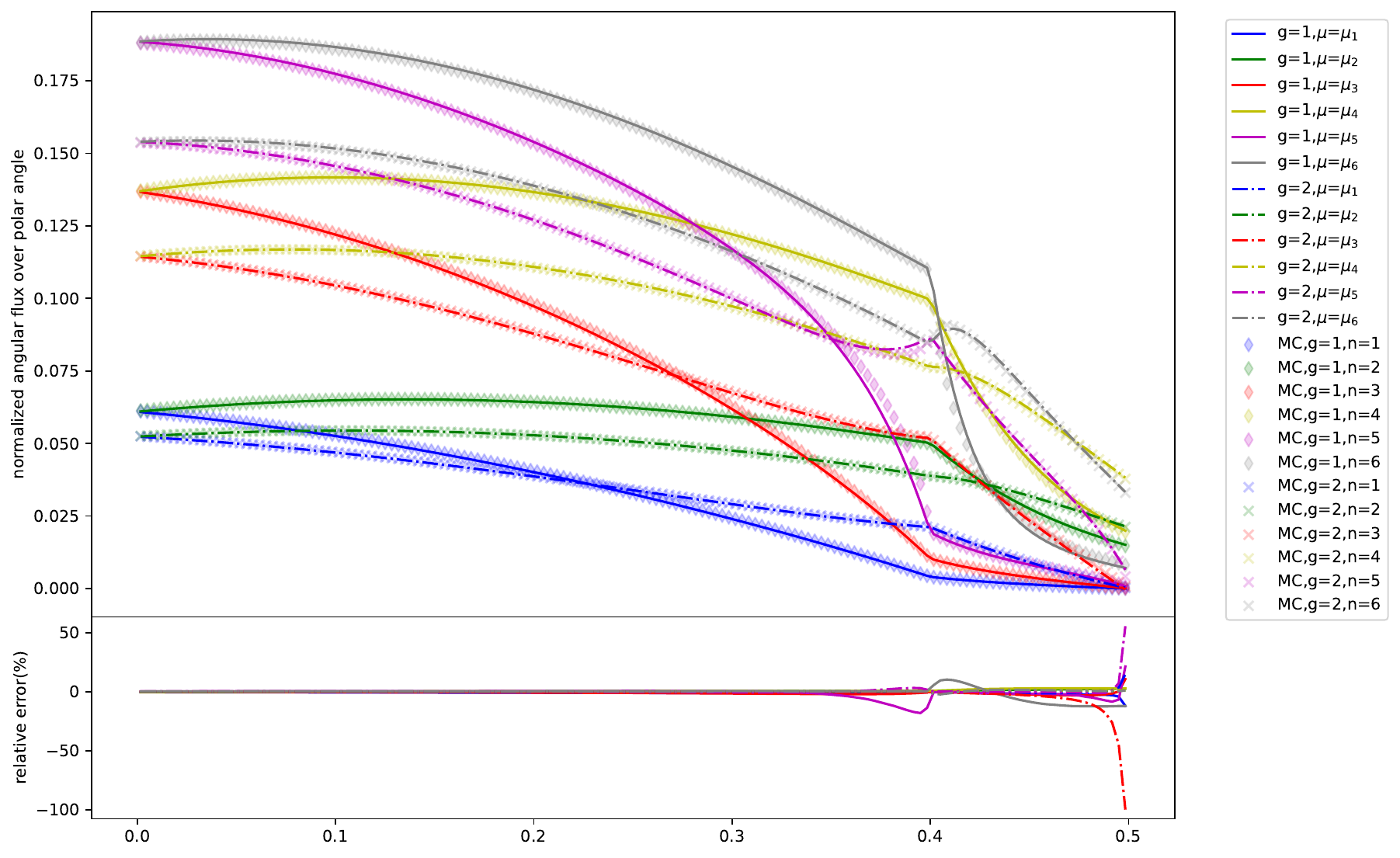}
\label{fig::psiS6}}  
\caption{Angular fluxes from $S_N$ ($w_n \psi_{g,n}$) vs Monte Carlo }
\label{fig::psi}
\end{figure}

Next, the angular fluxes are compared.
The discrete angles are ordered so that $|\mu_{2i+1}|=\mu_{2i+2}, \mu_{2}>...>\mu_N$. 
In openMC, for each angle $\mu_n$, fluxes are tallied for the angle range whose cosine spans range $\omega_n$.
These partial fluxes are compared with $\omega_n \psi_{g,n}$ from S$_N$ solution in Fig~\ref{fig::psi}. 
The S$_N$ approximation approaches the Monte Carlo reference when discrete order increases 
from S$_2$ in Fig~\ref{fig::psiS2}, to S$_4$ in Fig~\ref{fig::psiS4}, to S$_6$ in Fig~\ref{fig::psiS6}.
In addition to the visually well matched partial flux shapes,
the relative error between S$_N$ and reference at each position is also plotted in the lower part for each figure. 
The relative error improves with discrete order except S$_6$ at the close to zero values at right boundary.

\section{Conclusions}
In this work, we developed the method to solve multigroup S$_N$ equations in slab geometry. 
The method is accurate because it does not assume solution form like linear in fine meshes or polynomial in coarser meshes.
The method is efficient because the mesh can be as coarse as the material regions. 
The theory was verified with Monte Carlo simulations. 

In future effort, we will extend the method to cases with external sources. 
When the 1D S$_N$ solver is applied as axial solver in 2D-1D schemes, the radial traverse leakage is viewed as external source for the 1D problem. 
The current solution requires to inverse a sparse $NGR\times NGR$ matrix (fixed source problem) or solve non-linear equation to make its determinant 0 (eigenvalue problem). 
We will also explore performance improvement options.


\section{Acknowledgments}
This work is supported by the Department of Nuclear Engineering, The Pennsylvania State University.

\begingroup
\scriptsize
\bibliographystyle{ans}
\bibliography{SN1D}    
\endgroup
\end{document}